\documentclass[11pt,notoc]{JHEP3}

\usepackage{epsfig}
\usepackage{graphicx}
\usepackage{amssymb}
\usepackage{amsmath}
\usepackage{latexsym}

\renewcommand{\d}{\delta}

\renewcommand{\d}{\delta}
\newcommand{\be}{\begin{eqnarray}}
\newcommand{\ee}{\end{eqnarray}}
\newcommand{\del}{\partial}

\title{The Plasma Puddle as a Perturbative Black Hole}
\author{Clifford Cheung and Jared Kaplan \\
Jefferson Physical Laboratory, Harvard University, Cambridge, MA
02138 \\ E-mail: \email{cwcheung@fas.harvard.edu},
\email{kaplan@physics.harvard.edu} }

\abstract{We argue that the weak coupling regime of a large $N$
gauge theory in the Higgs phase contains black hole-like objects.
These so-called ``plasma puddles'' are meta-stable lumps of hot
plasma lying in locally un-Higgsed regions of space.  They decay
via $\mathcal{O}(1/N)$ thermal radiation and, perhaps
surprisingly, absorb all incident matter.
We show that an incident particle of energy $E$ striking the
plasma puddle will shower into an enormous number of decay
products whose multiplicity grows linearly with $E$, and whose
average energy is independent of $E$.
Once these ultra-soft particles reach the interior they are
thermalized by the plasma within, and so the object appears
``black.'' We determine some gross properties like the size and
temperature of the the plasma puddle in terms of fundamental
parameters in the gauge theory. Interestingly, demanding that the
plasma puddle emit thermal Hawking radiation implies that the
object is black (i$.$e$.$ absorbs all incident particles), which
implies classical stability, which implies satisfaction of the
Bekenstein entropy bound. Because of the AdS/CFT duality and the
many similarities between plasma puddles and black holes, we
conjecture that black objects are a robust feature of quantum
gravity.}

\begin{document}

\section{Introduction}

The AdS/CFT correspondence \cite{Maldacena:1997re}, \cite{Witten:1998qj}
\cite{Gubser:1998bc} has greatly
improved our understanding of both gravity and gauge theory by
providing a concrete realization of the holographic principle. For
example, much work has been devoted to studying strongly coupled
quasi-CFT dynamics using perturbative gravity. Conversely, CFTs have
been useful for illuminating aspects of black hole physics,
including the unitarity of Hawking evaporation.

It has been argued \cite{Aharony:2005bm} that there exist black
holes that can be localized in the IR of asymptotically AdS
geometries, and that these solutions are dual to ``plasma balls'' in
a confining CFT (for related work, see \cite{Giddings:2002cd},
\cite{Kang:2004jd}, \cite{Nastase:2005rp}). These plasma balls are
meta-stable lumps of hot gluon plasma, and like black holes, they absorb
all incoming matter and radiate thermally.
Interestingly, since the AdS/CFT duality maps quantum effects to
classical effects and vice versa, Hawking radiation is nontrivial
from the gravitational point of view but straightforward in terms of
the dual plasma ball description. Conversely, the ``blackness'' of
black holes -- their ability to absorb all incoming particles -- is
not obvious in the confining CFT picture.

In particular, consider the CFT dual of a particle thrown into a
black hole: a glue ball thrown into a plasma ball. Naively, it seems
that with sufficiently high energy such a glue ball would blast
through, in the same way that an extremely high energy proton might
barrel through the RHIC fireball. From this point of view, the most
fundamental property of black holes -- that they absorb all incoming
particles -- appears to be violated.

In \cite{Aharony:2005bm}, this problem was beautifully solved by
taking into account the parton substructure of the incident
particles \cite{Polchinski:2002jw}, in accordance with Susskind's
ideas \cite{Susskind:1994hb,Susskind:1994vu}. In the dual gauge
theory, the only available objects outside the plasma ball are
mesons and glue balls, and at large 't Hooft coupling these highly
boosted hadrons contain a huge number of soft partons.  Thus it is
simply impossible to fire a high energy parton into the plasma ball
-- instead, an incoming glue ball fragments into many low energy
partons which are promptly absorbed.

The purpose of the present paper is to explore the possibility
that a \emph{weakly} coupled gauge theory might furnish a
perturbative dual to a black hole.  At first this might seem like
an unlikely prospect, particularly since the dual of a weakly
coupled CFT is a strongly coupled gravitational theory.  Indeed,
it is not unreasonable to expect a phase transition between the
weak and strong coupling regimes of the gravitational theory,
corresponding to immensely different CFT physics in the two
regimes. Despite these expectations, however, we will argue that
``plasma puddles'' in a weakly coupled gauge theory are
qualitatively very similar to plasma balls/black holes.

Our setup is similar to \cite{Aharony:2005bm}, except that we
consider a perturbative gauge theory in the Higgs phase, rather
than a strongly coupled gauge theory in the confining phase. The
low energy theory is comprised of photons and $W$ bosons, rather
than mesons and glue balls.  Specifically, we will study an
$\mathcal{N}=4$ $SU(N)$ SYM theory Higgsed down to $U(1)^{N-1}$.
By heating up a region of space we can locally un-Higgs the gauge
group (see figure 1), creating a spatially varying Higgs vev. This
means that $W$ bosons have a position dependent mass that will act
as an effective potential for the enclosed plasma.  We will see
that for sufficiently large puddles, the plasma has a temperature
much less than the height of the enclosing potential, and so it is
kinematically trapped.  Thus, classical stability is ensured. That
said, the plasma puddle does emit radiation in a thermal spectrum,
but we find that its lifetime is \be \tau \sim N R , \ee where $R$
% XXX
is the radius of the puddle, so plasma puddles are
infinitely long lived in the large $N$ (classical) limit.

Since our theory is weakly coupled, it is possible to analyze the
absorptive properties of a plasma puddle using standard perturbation
theory. Unlike glue balls, photons and W bosons have no partonic
substructure -- since they are elementary point particles,
arbitrarily large boosts involve no partonic subtleties.  Thus it
would seem that at sufficiently high energies, a plasma puddle can
be penetrated. However, incident particles actually experience
something very similar to parton showering -- they decay near the
boundary of the plasma puddle, bifurcating into a large number of
ultra-soft daughter particles whose ensemble energy is smeared over
the surface of the puddle and eventually thermalized. Qualitatively, this matches
the process of gravitational ``hair removal'' in which an infalling
particle is delocalized over the surface of a black hole as it is
absorbed. In fact, this showering can only occur at the plasma
puddle boundary (the effect requires momentum nonconservation),
which dovetails nicely with the notion that black hole absorption is
a local effect at the event horizon, independent of the interior.

\begin{figure}[h]
\begin{center} \label{FigPlasmaPuddle}
\includegraphics[width=14cm]{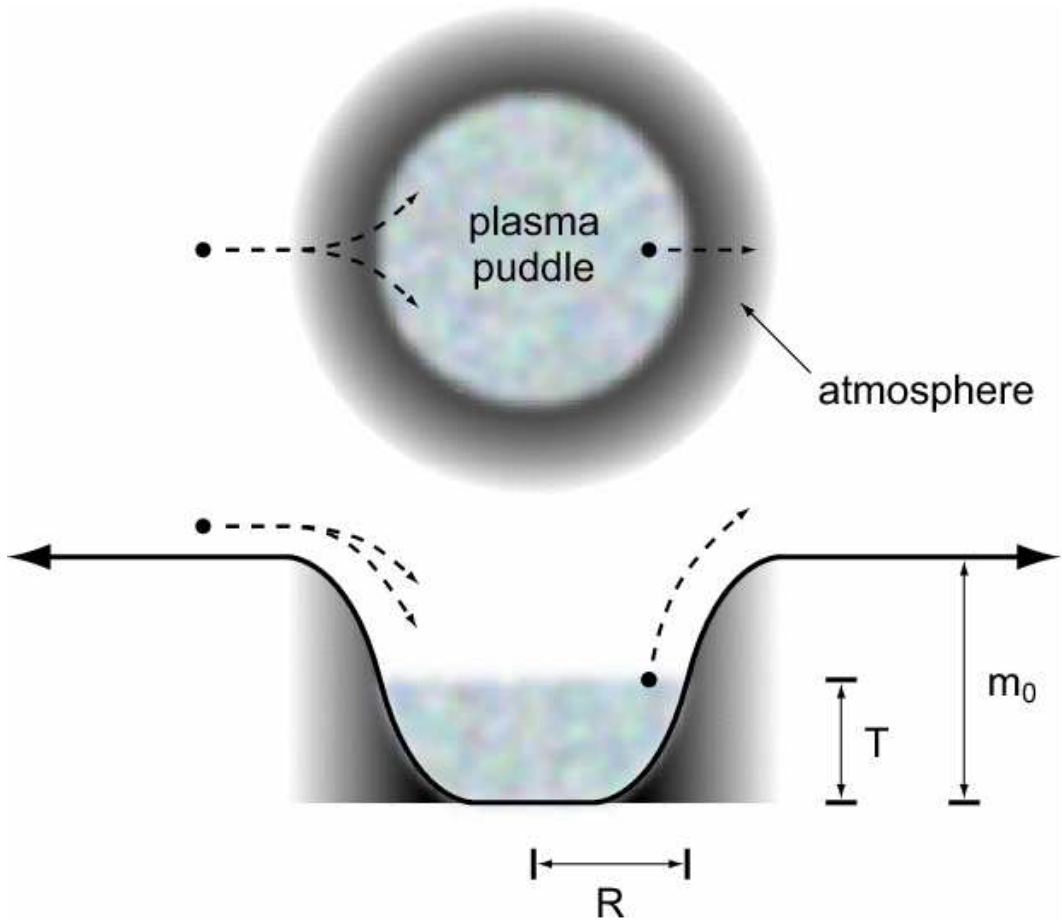}
\caption{\small{A plasma puddle cross section (above) and energy
profile (below).  The latter depicts the $W$ mass as it asymptotes
to a value of $m_0$ at infinity and vanishes within an un-Higgsed
region of radius $R$. The plasma within settles into a hot puddle
of temperature $T$. Both diagrams show a high energy incident
particle on the left showering in the atmosphere, as well as some
Hawking radiation escaping on the right.}}
\end{center}
\end{figure}

In order for a plasma puddle to absorb even the highest energy
particles, an incident particle of energy $E$ has to shower into a
large number of decay products that are too soft to escape the
enclosing potential. In particular, it is necessary that the
average energy of the final decay products, $E_{\rm avg}$, does
not increase with increasing $E$. If this is not the case, then an
arbitrarily large $E$ implies a commensurately large $E_{\rm
avg}$, allowing the decay products to blast through the plasma
unharmed. Thus, total absorption is only possible if the
% XXX
rate\footnote{For a detailed calculation of the dimensionless
\emph{probability} of showering, see appendix
\ref{sec:crosssection}} of showering increases at least linearly
with energy.  For $E$ less than a particular (large) threshold we
show that this is the case: the rate of decay of a gauge boson to
two $W$'s is \be \Gamma_{A \rightarrow W W} &\sim& \lambda E,
\label{tHans} \ee where $\lambda$ is the 't Hooft coupling.
Moreover, even though the decay rate goes to a constant for
incident particles above this threshold energy, we find that they
are still absorbed.  In particular, these high energy particles
shower promptly inside the plasma puddle into a large number of
decay products at the threshold energy, which in turn decay at a
rate $\lambda E$. Thus, a high energy incident particle will
shower into $\mathcal{O}(\lambda ER)$ decay products each of
energy $E_{\rm avg} \sim (\lambda R)^{-1}$. We confirm this
reasoning with a more precise argument in section
\ref{sec:absorption}.  As long as $E_{\rm avg}$ is less than the
height of the enclosing potential, these decay products are
trapped inside the plasma puddle, and so the puddle appears black.

Interestingly, as we show in section \ref{sec:implications}, if we
demand that the plasma puddle emit thermal Hawking radiation, then
it \emph{automatically} also absorbs all incident matter, is
classically stable, and satisfies the Bekenstein entropy bound
\cite{Bekenstein:1980jp}.  During the process of Hawking
evaporation, a stable plasma puddle will lose each of these
properties sequentially, until it eventually becomes a free gas of
gauge bosons.  Since the plasma puddle is so similar to a black
hole, our hope is that the large $\lambda$ plasma ball/black hole
duality established in \cite{Aharony:2005bm} interpolates at small
$\lambda$ to a correspondence between the plasma puddle and some
black object of a strongly coupled gravitational theory.

Note that for our purposes we will focus entirely on the gauge
degrees of freedom of the $\mathcal{N}=4$ SYM, ignoring gaugino
partners and scalar moduli. Moreover, as we are interested only in
parametric scalings, we will be largely ignoring numerical
factors. The outline of the paper is as follows.  In section
\ref{sec:general} we determine the general properties of our
setup: gross properties of the plasma puddle, Hawking radiation,
and classical stability.  In section \ref{sec:gauge}, we give a
nice physical estimate of the rate of particle showering in the
plasma puddle atmosphere. In section \ref{sec:implications} we
show that demanding thermal Hawking radiation immediately implies
other black hole-like properties, and we conclude in section
\ref{sec:discussion}.  The appendices contain a more formal
derivation of the probability of decay.

\section{The Plasma Puddle}

\label{sec:general}

In this section we give a detailed account of what a plasma puddle
is and how it forms.  Our setup is as follows. %Consider an $\AdSS$
%geometry containing a stack of $N$ noncoincident d-branes. A
%collection of open strings on these d-branes will naturally collapse
%and coalesce into a hot region of space, yielding a localized black
%hole.
Consider an $\mathcal{N}=4$ $SU(N)$ SYM at large $N$ and weak 't
Hooft coupling $\lambda$ that is Higgsed down to $U(1)^{N-1}$
(note that due to the Higgsing, the theory is not conformal and so
particles and S-matrices are well defined). The spectrum of the
theory consists of $N-1$ massless photons and $N^2 - N$ massive
$W$ bosons. Now, let us fire an ensemble of high energy $W$'s into
a small region of space\footnote{Note that an ensemble of $W$
bosons that are initially at rest will naturally collapse due to
the attractive force of dilaton gravity, as in described in
\cite{Sundrum:2003yt}. In the nonrelativistic limit, this behaves
exactly like true gravitational collapse.}. The influx of $W$'s
heats up the region and locally un-Higgses the gauge group. Once
the $W$'s thermalize, the resulting meta-stable object is a plasma
puddle.

 The local un-Higgsing can be parameterized by a
spatially varying Higgs vev which induces a spatially varying
mass, $m(x)$, for the $W$ bosons.  The $W$ mass profile vanishes
inside the plasma puddle but asymptotes to some nonzero mass $m_0$
outside (the precise mass of individual $W$s depends on the
Higgsing pattern, but the details will not be important). Let us
define the atmosphere to be the region near the plasma puddle
boundary in which the mass profile varies.  Again, we emphasize
that the $W$'s are confined by the $m(x)$ potential, while the
photons are actually massless everywhere.

If we ignore all interactions, it is straightforward to see what
happens to free streaming $W$'s as they collapse into a plasma
puddle. Those $W$'s with $E>m_0$ escape to infinity while those with
$E<m_0$ settle at the basin of the $m(x)$ potential.  Due to the
potential barrier, this puddle of $W$'s can never escape.

The story is similar if we include gauge interactions, except that
the puddle of $W$'s thermalizes into a puddle of photons and $W$
bosons. To see this, we compute the mean free path $d$ of a gauge
boson $A_{ij}$ traversing the hot plasma. From color conservation,
$A_{ij}$ can only scatter off of some $A_{jk}$, leaving $A_{il}$ and
$A_{lk}$ in the final state, where $k$ and $l$ are free indices.
Summing the amplitude squared, $g^4$, over phase space, $k$ and $l$
contribute $N^2$ to the cross section, yielding \be \sigma &\sim&
\lambda^2 T^{-2}, \ee since the temperature sets the energy scale of
the interaction.  The number density $n$ of a gauge boson with a
given pair of color indices is $T^3$, so the mean free path is \be d
&\sim& (n \sigma)^{-1} \sim (\lambda^2 T)^{-1} . \ee

For sufficiently large 't Hooft coupling the mean free path is
smaller than the size of the plasma puddle and thus gauge bosons
cannot go very far without scattering (see figure 2).  Let us denote
this as the highly thermalized regime. In this case the plasma
puddle quickly thermalizes into a hot, homogenous soup of photons
and $W$ bosons at a temperature $T$.  The plasma puddle is
classically stable as long as its temperature is less than $m_0$,
so that the interior plasma is kinematically
trapped. Without this condition, nothing prevents the plasma from
simply escaping to infinity, and so the relation \be T < m_0
\label{eq:classstab},\ee implies classical stability of the plasma
puddle. In addition, since the plasma puddle is highly thermalized,
only particles at the very surface have any hope of escape. We argue
in section \ref{sec:hawking} that this phenomenon is an
$\mathcal{O}(1/N)$ area effect that is thermal and dual to Hawking
radiation. For these reasons (along with their absorptive properties,
which we discuss in section \ref{sec:gauge}) we claim that plasma puddles in the
$d<R$ regime are black hole-like objects.

On the other hand, if $\lambda$ is sufficiently small, then  $d > R$
and a typical gauge boson can traverse the extent of the plasma
puddle without ever scattering  (see figure 2). However, given a
time of order $d$, the plasma will eventually thermalize, yielding a
collection of nearly free photons and $W$'s at a temperature $T$.
The $W$ bosons travel in straight lines through the puddle until
they reach the potential barrier from $m(x)$, after which they roll
back towards the interior, and repeat. As a result the $W$'s form a
relativistic plasma at the bottom of the $m(x)$ potential, but they
become nonrelativistic in near the $m(x)$ barrier wall, simply
because they have less kinetic energy there. In contrast, the
photons are massless everywhere and can free stream outwards.  As
we discuss in section \ref{sec:hawking}, since plasma puddles in the $d
> R$ regime do not Hawking radiate in the traditional sense, we do
not identify them as dual black holes.

\subsection{Gross Properties and Relations}

\label{sec:grossprop}

Thus far our discussion has included $T$ and $R$ as a priori
attributes of the plasma puddle. However, as we show in this
section, these two variables are fixed in terms of the ``universal''
quantities $N$, $\lambda$, $m_0$, and the total energy of the plasma
puddle, $M$.\footnote{Here we use the symbol $M$ in anticipation of
matching this total energy of the plasma puddle to the mass of a
dual black hole.}

To begin we note that unlike Schwarzchild black holes, plasma
balls/puddles do not obey the relation $T \sim 1/R$ for the
following reason. Since the plasma puddle has an entropy $S = N^2
T^3 R^3$, the relation $T \sim 1/R$ would imply that the entropy is
independent of the size of the object. However, this is not the case
-- the entropy of a plasma puddle/ball increases with size.
In the case of the strongly coupled plasma balls of
\cite{Aharony:2005bm}, the temperature of large plasma balls is set
by the confinement scale. Analogously, one might expect a similar
situation for plasma puddles, i$.$e$.$ that the temperature is given
by the Higgs vev. We will see that this is not the case.

Once formed, a plasma puddle is an isolated system, so its total
energy is conserved.  However, its entropy should be maximized
subject to this constraint, so we treat the plasma puddle with the
micro-canonical ensemble. In $\mathcal{N}=4$ SYM, the scalar fields
$\Phi_{ij}$ are in the adjoint representation of the gauge group.
When these fields Higgs $SU(N)$ to $U(1)^{N-1}$, their vevs can be
written as diagonal $N\times N$ matrices with spatially varying
components. The total energy of a plasma puddle arises from two
contributions: the gradient kinetic energy from the (spatially
varying) Higgs mechanism, and the thermal energy of the plasma
within. Because these contributions are boundary and volume effects,
respectively, we are essentially balancing the pressure against the
surface tension. Neglecting $\mathcal{O}(1)$ factors, the total
energy of the plasma puddle is thus \be M = \int d^3 x \, {\rm Tr}
(\nabla \Phi)^2 + N^2 T^4. \ee The second term only accounts for the
relativistic ($T \gg m(x)$) region of the plasma. The
nonrelativistic ($T \ll m(x)$) region has an energy density $N^2 T^2
m(x)^2 \exp(-m(x)/T)$, and this is always less than or equal to $N^2
T^4$.  Thus neglecting this contribution merely amounts to a
rescaling of the plasma energy by an $\mathcal{O}(1)$ factor.

 Each diagonal entry of $\Phi$ gets a vev that asymptotes to $\sim m_0 / g$
at infinity but vanishes in some region of size $R$. Moreover, let
us define $L$ to be the thickness of the atmosphere, i$.$e$.$ the
region in which the Higgs vev varies.  Estimating the energy we find
that \be M = N^2 \left( \frac{m_0^2 (R+L)^2}{\lambda L} +  T^4 R^3
\right). \ee

Before maximizing the entropy at fixed energy, let us motivate
why this fixes the radius of the plasma puddle.  Consider what
happens as we increase the radius of the puddle at fixed
$m_0$.  This increases the gradient $\Phi$ energy, but since the
total energy is fixed, this forces the temperature to diminish.
Since the entropy goes as $T^3 R^3$, the increase in radius and
decrease in temperature are competing effects.  Hence, there is
an extremal value for $R$ such that the entropy is maximized.  Note
that this is what fixes the size of a balloon full of gas.

Fixing $M$ allows us to solve for $T$, yielding an entropy \be S &=&
N^2 T^3 R^3 \\  &=& \sqrt{N} R^{3/4} \left( M - \frac{N^2 m_0^2
(R+L)^2}{\lambda L} \right)^{3/4}. \ee The entropy is maximized for
$L \sim R$, because this maximizes the term in large parentheses.
Maximizing the entropy with respect to $R$, we find that \be R \sim
\frac{\lambda M}{N^2 m_0^2}, \ee and also that
\be T &\sim& \frac{N m_0^{3/2}}{\lambda^{3/4} M^{1/2}}.
\ee Thus we see that while strongly coupled plasma balls obey the
relation $R \sim M^{1/3}$ \cite{Aharony:2005bm}, perturbative plasma
puddles have that $R \sim M$.

Finally, let us add some remarks about the precise shape of the
Higgs vev, $\Phi$.  Near the boundary of the plasma puddle
it is nontrivial to determine $\Phi$,
because the plasma back-reacts on the Higgs vev and vice
versa. However, outside $R$ the $\Phi$ field is free, so it obeys
Laplace's equation, and thus \be \Phi \sim \left( 1 - \frac{R}{r}
\right) \quad \mathrm{for} \quad r
> R, \ee obtains. Thus the scalar field atmosphere has a $1/r$ tail,
so again we see that the thickness $L$ of the plasma puddle is of
order $R$. In $d \geq 4$ spacetime dimensions (in the gauge theory),
the Higgs profile would go as \be \Phi \sim 1 - \left(\frac{R}{r}
\right)^{d-3}, \ee so in higher dimensions there are no IR
subtleties.  For this reason we do not expect that this tail is
qualitatively important for black hole/plasma puddle physics.

\subsection{Hawking Radiation}

\label{sec:hawking}

In the gravitational picture, the total absorption of incident
particles by a black hole is obvious at the classical level, while
Hawking radiation is a nontrivial quantum effect. In contrast, in
our CFT setup we will see that plasma puddle absorption is
nontrivial but Hawking radiation is manifest. In this section we
consider the latter.

To begin, let us consider the regime in which $d < R$ and the
enclosed plasma is highly thermalized.  In this case the gauge
bosons cannot go very far without scattering, and thus a particle
has no hope of escape unless it is at the very surface of the
plasma puddle. Moreover, since the $m(x)$ potential kinematically
bounds $W$ bosons, a surface particle can only escape if it is a
photon. Since only $1/N$ of the particles is a photon (due to the
Higgsing pattern $SU(N)\rightarrow U(1)^{N-1}$), the plasma puddle
radiates photons in a thermal spectrum at temperature $T$. Since
this is a $\mathcal{O}(1/N)$ effect, it is natural to associate
this radiation to $\mathcal{O}(\hbar)$ Hawking radiation in a
gravitational dual.

Next, let us compute the rate at which a plasma puddle loses energy
as the result of Hawking radiation.
In an infinitesimal time interval $dt$, an order one fraction of
photons that are within $dt$ of the surface of the plasma puddle
will exit. If $n_\gamma$ is the number density of photons in the
plasma, then there are $n_\gamma R^2 dt  $ photons in this region
near the surface.  Since each has an energy of order $T$, \be
\frac{d E}{dt} &=& - n_\gamma R^2 T  . \ee Assuming that these
photons are in thermal equilibrium, we find
\be \label{PowerRadiated} \frac{d E}{dt} &=& - N R^2 T^4 \nonumber \\
&=& -\frac{N m_0^2}{\lambda}, \ee so we see that at fixed
temperature the energy loss is an area effect, while at fixed
$m_0$ the energy loss is completely independent of size and
temperature. The lifetime is given by the time it takes for the
total energy $M$ to be radiated \be \tau = -M \left( \frac{dE}{dt}
\right)^{-1} \sim N R , \ee so plasma puddles are completely
stable in the classical limit $N \rightarrow \infty$.

The situation is more complicated if $d>R$, so let us be more
careful. We consider scattering inside the plasma puddle for
arbitrary $d$ and $R$. If the overall number density of particles
is $n_{\rm tot}$, then there is a total of $n_{\rm tot} R^3$
particles, and so the total rate of scattering processes is
$n_{\rm tot} R^3 /d$. Of these processes, an $\mathcal{O}(1/N)$
fraction create additional photons. Likewise, there are $ n_\gamma
R^3$ photons in the plasma, so the rate for scattering events
involving photons is $n_\gamma R^3 / d$. Almost all of these
scattering events destroy photons, because $A W \rightarrow A W$
is suppressed by a factor $N$ compared to $A W \rightarrow W W$.
Thus scattering events change the number of photons in the plasma
at a rate \be \left. \frac{d ( n_\gamma R^3 ) }{dt} \right|_{\rm
scattering} = \left(\frac{1}{N} n_{\rm tot}  - n_\gamma \right)R^3
\frac{1}{d}. \ee However, radiation leaving the plasma puddle also
decreases the number of photons at the rate \be \left. \frac{d (
n_\gamma R^3 ) }{dt} \right|_{\rm radiation} = - n_\gamma R^2  .
\ee These two processes balance when \be n_\gamma \sim \frac{1}{N}
n_{\rm tot} \left(1+\frac{d}{R}\right)^{-1}.\ee Thus we see that
as $d$ becomes larger than $R$, the number of photons becomes less
than $1/N$ of the total number of particles, and so we leave
thermal equilibrium. In particular, once $d > R$, the rate of
Hawking radiation begins to decrease by the significant factor
$R/d$. Thus, in the $d > R$ regime the plasma puddle does not
radiate a thermal spectrum, and exiting photons can free stream
from anywhere within the plasma, not just the surface.  For this
reason we do not consider such a plasma puddle to be the dual of a
black hole.

\begin{figure}[th]
\begin{center}
\includegraphics[width=14cm]{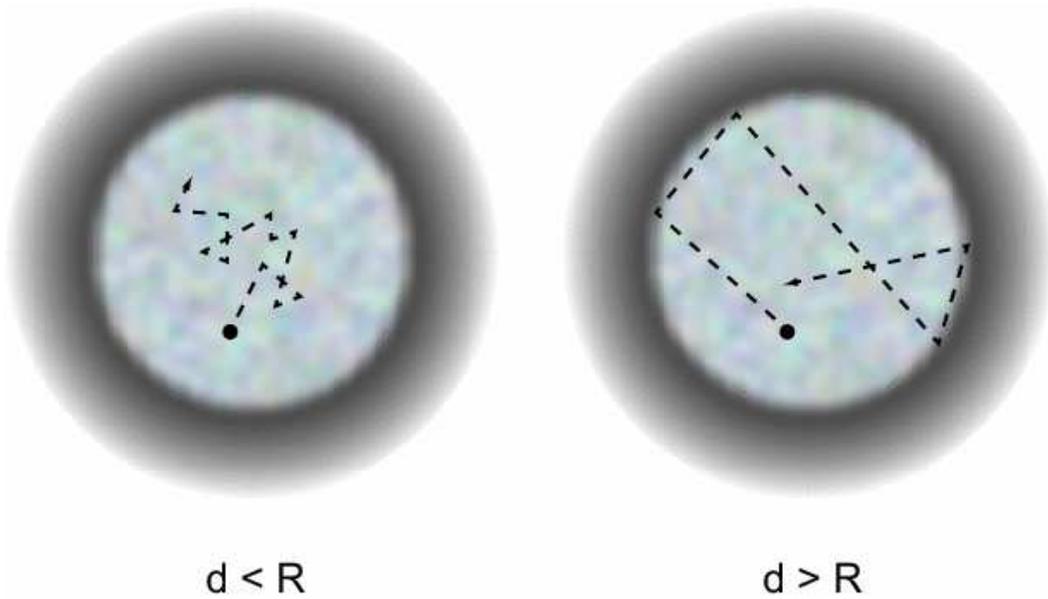}
\caption{\small{Here we show the typical trajectory of a gauge boson
in the highly thermalized regime, $d < R$, and otherwise. Hawking
radiation is only thermal when $d < R$.}}
\end{center}
\end{figure}

We have glossed over a subtlety in our treatment of Hawking
radiation. As we argue in section \ref{sec:gauge}, a high energy
photon striking the plasma puddle atmosphere will shower into many
other gauge bosons, until all of the decay products have a tiny
energy of order $(\lambda R)^{-1}$.  But since it is possible that
$T > (\lambda R)^{-1}$ (in fact, an even stronger condition is
required to be in the highly thermalized regime), we might worry that
an outgoing photon will simply shower in the plasma puddle
atmosphere. If this happens, its low energy decay products are
likely to be reabsorbed by the plasma puddle, leaving no reason to
expect Hawking radiation, let alone a thermal distribution of
Hawking radiation.

The resolution to this puzzle is that once an outgoing photon makes
it to the outer atmosphere of the plasma, it does not have the $2
m(x)$ of energy necessary to shower into two $W$ bosons, since it
only has energy $T$. Thus, kinematical constraints allow photons at
temperature $T$ to escape from the plasma puddle as thermal Hawking
radiation, even though high energy incident photons typically shower
as they fall in.

Finally, let us discuss briefly the possibility of $W$ boson Hawking
radiation.  While most of the $W$ bosons are of course kinematically
trapped, there is a Boltzmann tail in the thermal distribution which
allows an $e^{-m(x)/T}$ fraction of the $W$'s to escape.  While this
effect is exponentially suppressed, it is not suppressed by powers
of $1/N$.  This is puzzling from the standpoint of the gravity dual,
in which all Hawking radiation should be an $\mathcal{O}(\hbar)$
effect.  Understanding the true $N$ dependence of $W$ boson Hawking
radiation will require further study.

\begin{figure}[h]
\begin{center}
\includegraphics[width=14cm]{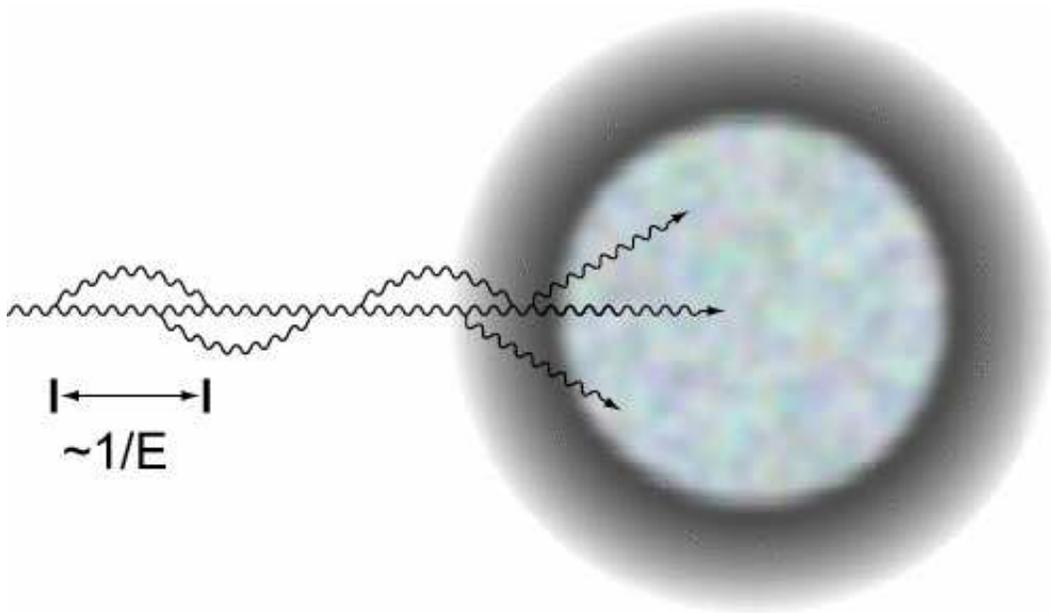}
\caption{\small{An incoming gauge boson is surrounded by a cloud of
virtual particles.  Only when the gauge boson strikes the plasma
puddle atmosphere can these virtual particles become real.  The
characteristic length scale of quantum fluctuations is given by
$(\lambda E)^{-1}$. }}
\end{center}
\end{figure}

\section{Plasma Puddle Absorption}

\label{sec:gauge}

The primary feature of the plasma puddle that distinguishes it
from more conventional objects that are hot and stable (such as
stars) is its ability to absorb all incident matter.  In this
section we verify that the plasma puddle is black by showing that
a high energy incident particle will shower into numerous soft
decay products that thermalize with the interior plasma.  While a
formal calculation of the decay probability is given in appendix
\ref{sec:toy}, we present a more direct physical argument in the
following section.

\subsection{Estimation of the Showering Rate}

To begin, let us consider the propagation of a single photon
through free space. The photon is absolutely stable due to
kinematics and phase space -- momentum conservation only allows
decays into exactly collinear, massless decay products and since
this is a set of measure zero in phase space the decay rate
vanishes.  However, in the presence of momentum violation (such as
the plasma puddle atmosphere), noncollinearities are allowed and
so the decay rate is nonzero.

 To estimate this rate, we
invoke the uncertainty principle, which tells us that a photon in
the vacuum is immersed in a $\mathcal{O}(\lambda)$ cloud of
virtual gauge bosons constantly coming in and out of existence.
Since the only relevant time scale for a massless particle in free
space is its energy $E$, the photon produces virtual particles of
energy $\sim E$ at a rate $\Gamma \sim \lambda E$.  Indeed, once
the photon strikes the plasma puddle atmosphere, the resulting
momentum violation allows for these virtual particles to become
real\footnote{This is a common effect, and is detailed in standard
textbooks on quantum field theory \cite{Peskin}. For example,
consider the electron, which carries with it a number density of
virtual photons given by $N_\gamma \sim \alpha$. Bremsstrahlung
effects from the scattering of an electron off of a charged target
can be understood as virtual photons becoming real due to a
nontrivial background.}.
% XXX
Since $\Gamma$ is the rate at which photons are produced, and in the
momentum violating background they can simply escape to infinity, we expect
that \be \Gamma_{A\rightarrow WW} \sim \lambda E . \ee
% we can simply
% multiply by $R$ to obtain the (dimensionless) probability of decay
% across the extent of the plasma puddle.  Hence, an infalling particle
% will decay to $W$'s with a probability \be P_{A\rightarrow WW} \sim
% \lambda E R, \ee in the momentum violating background.  Note that this
% probability can be greater than one, but this simply indicates that
% the decay rate is large enough that the particle could decay more
% than once as it traverses the plasma puddle.

The reader may suspect that this answer breaks down at large energy,
and this is in fact the case.  There are several reasons why we
might expect this to happen.  First of all, if the photon is
arbitrarily boosted (relative to the rest frame of the plasma
puddle), it eventually reaches the threshold energy for plasma
puddle creation, and so the correct calculation would involve
plasma puddle/plasma puddle scattering.  This is akin to
firing a particle into a black hole at trans-Planckian energies,
which is really the same as black hole/black hole scattering.

Secondly, at very high energies the $W$ decay products of the photon
are so collinear that their splitting is completely unmeasurable (by
the plasma puddle) and is thus unphysical.  Let us illustrate this
soft collinear divergence more explicitly.
% XXX
As noted previously, the photon cannot decay into two exactly
collinear $W$ bosons due to phase space, but if it receives of a
momentum ``kick'', then it can decay into $W$s which split in the
transverse direction.  However, it can only decay in this way in
the momentum violating background. In the plasma puddle
atmosphere, where the plasma itself is negligible, the only source
of momentum violation (position dependence) is the gradient of the
Higgs field, so we expect that the kick is proportional to $1/R$.
The angle subtended by the outgoing $W$'s is approximately
$k_\perp /E$, where $k_\perp^2 \sim E /R $ is the transverse
momentum difference between the $W$'s\footnote{We estimate
$k_\perp$ as follows.  We begin with an incident photon with
4-momentum $(E, 0,0,E)$ and assume that it receives a momentum
``kick'' $(0,0,0,-1/R)$. This allows it to split into two massless
particles with $k_\perp^2 \approx E^2 - (E - 1/R)^2 \sim E / R$}.
Demanding that the ``cone'' traced out by the outgoing particles
grows to a size greater than their Compton wavelength $\sim
1/m_0$, we find that \be R \frac{k_\perp}{E} > \frac{1}{m_0} \quad
\Rightarrow \quad E < m_0^2 R . \quad (\textrm{soft collinear
bound}) \label{eq:softcoll}\ee If this bound is violated, then
there is a soft collinear divergence and the splitting is
unmeasurable. Note that if the energies of the decay products are
different, only the lesser of the two need satisfy this bound.

In appendix \ref{FromToyScalarstoGaugeBosons} we address these
high energy issues directly by calculating the rate of decay
across the extent of the plasma puddle, yielding
 \be \Gamma_{A \rightarrow W W} &\sim& \lambda E, \quad (E <
 m_0^2 R) \\ &\sim& \lambda m_0^2 R, \quad (E >
 m_0^2 R, \textrm{Bremsstrahlung}) \\ &\sim& \frac{\lambda m_0^4 R^2}{E}, \quad (E
 >  m_0^2 R, \textrm{symmetric decay}), \ee where ``Bremsstrahlung'' denotes the
 region in phase space in which one of the outgoing particles is much softer
 than the other (that is, with energies of order $m_0^2 R$ and $E-m_0^2 R$) and ``symmetric decay'' denotes
 the regime in which they are comparably energetic (with energies of order $E$).  Since the probability of a
 symmetric decay is suppressed by a factor of $m_0^2 R /E$ relative to
Bremsstrahlung, soft emission totally dominates showering at
energies
 above the threshold $m_0^2 R$.  For example, consider
 an incoming particle boosted to some high energy a hundred times greater than
 $m_0^2 R$.  Emission of a soft particle of energy $m_0^2 R$ is a hundred times
 more likely than decay into two particles of comparable energy.
 Moreover, once this Bremsstrahlung occurs, the harder particle only
 loses $m_0^2 R$ energy, and so a consecutive soft emission is
 still ninety nine times more likely than a symmetric decay, and so
 on.  Before long, the particle emits enough Bremstrahlung that its energy
drops to $m_0^2 R$ and its decay rate begins to scale linearly
with energy.  In the following section we consider the physical
consequences of these rates.

\subsection{Absorption and $E_{\rm avg}$}

\label{sec:absorption}

In section \ref{sec:gauge} we saw that $P_{A \rightarrow W W}$
increases linearly with $E$ up to the scale $m_0^2 R$, after which
it remains constant.  Next, we show that this implies that an
incident particle, no matter how energetic, will always be
absorbed by the plasma puddle.

To begin, let us consider showering in the $E < m_0^2 R$ regime.
An incident particle of energy $E$ will decay in the plasma puddle
atmosphere at a rate $\lambda E$. In turn, its decay products have
a smaller rate for decaying simply because they have less energy.
In fact, after a long sequence of decays the average energy of the
final decay products eventually diminishes to $E_{\rm avg} \sim
(\lambda R)^{-1}$. At this point the decay rate dips below $1/R$,
and the ultra-soft daughter particles can traverse the extent of
the plasma puddle without showering.  Thus, showering completely
terminates once the decay products reach an energy of $(\lambda
R)^{-1}$. Since the final decay products have a Compton wavelength
proportional to the radius of the plasma puddle, showering
effectively de-localizes the incident particle over the entire
puddle. This picture matches nicely with Susskind's ideas about
black hole absorption \cite{Susskind:1994hb,Susskind:1994vu}.

In addition, as long as this terminal energy is less than $m_0$, the
daughter particles are kinematically trapped by the Higgs profile,
and they will eventually be thermalized.  This translates into a
blackness bound \be \frac{1}{\lambda R} < m_0, \label{eq:blackbound}
\ee which must be satisfied if the plasma puddle is to absorb all
incident matter.

We now argue that the above conclusions are still valid even above
the threshold energy, $ m_0^2 R$. As we argued earlier, such a
high energy particle favors the emission of soft particles of
energy $m_0^2 R$.  However, because the decay rate for
Bremsstrahlung remains constant at high energies, one might wonder
whether a sufficiently energetic particle might blast through the
plasma puddle. A careful analysis shows this is not the case.

In particular, let us very roughly estimate the rate of $1
\rightarrow n$ showering for an incident particle of energy $E = n
m_0^2 R$, where $n$ is some larger integer. Noting that a $1
\rightarrow n$ decay includes a sequence of $i \rightarrow i+1$
sub-processes, and is often dominated by on-shell regions of phase
space, we can estimate the rate of $1 \rightarrow n$ decay by \be
\Gamma_{1\rightarrow n} &=& \frac{1}{R} P_{1\rightarrow 2} \times
P_{2\rightarrow 3} \times  \ldots \times P_{n-1 \rightarrow n}, \\
&=& \frac{1}{R} (\lambda m_0^2 R^2)^n, \ee where schematically $P
\sim \Gamma R$ (see appendix \ref{sec:toy}, where we calculate
these probabilities explicitly). As we show in section
\ref{sec:implications}, classical stability immediately implies
that the quantity in parentheses is greater than unity. Thus, we
argue that an incident particle of energy $n m_0^2 R$ will decay
into $n$ particles of energy $m_0^2 R$ within a region of size
$R$.  After this sequence of Bremsstrahlung events, the decay
products all have energies of order $m_0^2 R$, and so we can apply
our analysis for the $E < m_0^2 R$ regime to these decay products.
Since these particles decay to a multitude of ultra-soft particles
of energy $(\lambda R)^{-1}$, we find that total absorption
occurs, no matter what the incident energy.

Interestingly, we are finding that large multiplicity events
dominate at high energies, which is reminiscent of Hawking
evaporation -- after all, the probability that a large black hole
will decay to two particles is extremely small, because it is
suppressed by $e^{-S}$, while the decay to a huge number of soft
particles is virtually guaranteed. Thus the necessity of including
$1 \rightarrow n$ decays at high energies seems to indicate that
high energy particles carry a great deal of entropy.

\section{Implications of $d < R$}

\label{sec:implications}

In section \ref{sec:grossprop} we argued that in the highly
thermalized regime ($d < R$) the plasma puddle emits thermal Hawking
radiation at the surface. In this section we show that $d < R $ also
implies total absorption, which implies classical stability, which
implies satisfaction of the Bekenstein entropy bound. \footnote{The
Bekenstein entropy bound gives an upper bound on the entropy of a
system in terms of its gravitational mass $M$ and its size $R$:  $S <
MR$.} For this reason, we identify $d < R$ plasma puddles as black
hole-like objects.

Applying our relations from section \ref{sec:grossprop}, we can
rewrite the $d < R$ bound, the blackness bound
(Eq.~(\ref{eq:blackbound})), the classical stability bound
(Eq.~(\ref{eq:classstab})), and the Bekenstein bound as
\be m_0 R &>& \lambda^{-7/2}, \quad (\textrm{highly thermalized}) \\
&>& \lambda^{-1}, \quad (\textrm{total absorption}) \\&>&
\lambda^{-1/2}, \quad (\textrm{classical stability})\\
&>& \lambda^{1/2}, \quad (\textrm{Bekenstein bound}).\ee The above
sequence of bounds gives an interesting depiction of what happens
during plasma puddle evaporation.  As the energy of the plasma
puddle diminishes via Hawking radiation, its size shrinks
commensurately. First, the plasma puddle leaves the highly
thermalized regime and stops emitting thermal Hawking radiation.
Next, the plasma puddle becomes transparent to incident matter, and
finally the object becomes classically unstable. After this point,
the enclosed photons and $W$ bosons in the plasma simply stream
outwards, leaving a free gas of gauge bosons.  It would be interesting
to explicitly connect these processes with the Horowitz-Polchinski
transition \cite{Horowitz:1996nw}, \cite{Alvarez-Gaume:2005fv},
\cite{Alvarez-Gaume:2006jg}.

It is also noteworthy that the Bekenstein bound is precisely
equivalent to the condition \be T > \frac{1}{R}, \ee which is
necessary for the application of classical thermodynamics.  As this
bound is approached, the energy/size of the system begins to
saturate the uncertainty principle, and we are forced to count
individual quantum mechanical microstates.

\section{Discussion}

\label{sec:discussion}

In this paper we have argued for the existence of black hole-like
objects living in large $N$ gauge theories at weak 't Hooft
coupling. Since these theories are completely perturbative, we can
calculate much of the physics.  In particular, in the regime in
which the mean free path $d$ is smaller than $R$, we find that these
meta-stable puddles of plasma are classically stable and emit
radiation in a thermal spectrum. While these properties are of
course common to any hot star-like object, we moreover find that the
plasma puddle absorbs all incident matter, no matter how energetic.
This occurs because high energy particles invariably shower into
ultra-soft decay products that are kinematically bound by the
effective potential from the spatially varying Higgs vev.  All in
all, we find this to be compelling evidence that the plasma puddle
is dual to a black object in a strongly coupled gravitational
theory.

In addition, our work gives a particularly nice picture of plasma
puddle evaporation, which may be connected to the
Horowitz-Polchinski transition \cite{Horowitz:1996nw},
\cite{Alvarez-Gaume:2005fv}, \cite{Alvarez-Gaume:2006jg}.  Indeed
from section \ref{sec:implications} we saw that as the object
radiates away energy, it eventually leaves the $d < R$ regime and
stops emitting thermal Hawking radiation. After even more energy
loss, the plasma puddle stops being black, and eventually becomes
classically unstable. Interestingly, we also find that the $d < R$
bound gives a nice lower bound on the total energy of a plasma
puddle, given by \be M &
>&  \lambda^{-9/2} N ^2 m_0. \ee  Since a plasma puddle must
have at least this much energy to form a black hole-like object,
we might interpret the left-hand side as the CFT dual to the
Planck mass.

\section*{Acknowledgements}

We would like to thank Ofer Aharony and Nima Arkani-Hamed for
helpful discussions, and Toby Wiseman for correspondence.  We also
thank Jonathan Heckman for collaboration when this work was at an
early stage. Jared Kaplan is supported by a Hertz Foundation
Graduate Fellowship.

\appendix

\section{Detailed Absorption Computation}

\label{sec:toy}

In this section we give a detailed derivation of the probability for the
decay of a photon to two $W$ bosons as it passes through the plasma
puddle atmosphere.  Although we are formally calculating a \emph{probability},
we can re-interpret it as a decay rate via $\Gamma \sim P/R$.
To begin, we consider a toy scalar model for simplicity.  This
scalar field theory has an action \be S &=& \frac{1}{2} \int \del
\phi^2 + \del\chi^2 - (m^2(x_3) + g \phi) \chi^2 . \ee In this
theory, a massless scalar field $\phi$ is coupled cubically to a
scalar $\chi$ whose mass $m^2$ is a nontrivial function of $x_3$.
Here $\phi$ and $\chi$ are scalar analogs of the photon and the $W$
boson, respectively, and $m^2$ mimics the effects of a space varying
Higgs vev on the $W$ mass.  Notice that we have made the
simplification that the plasma puddle is infinite in the $x_1$ and
$x_2$ directions, which is a very good approximation if the plasma
puddle is large and the incoming particles approach from the $x_3$
direction.

\subsection{Propagator}

Next, let us compute the $\chi$ propagator in the regime in which
the incoming and outgoing energies are much larger than the
characteristic energy scale $1/R$ set by the mass profile.  To do this
we use the WKB approximation to solve the wave equation\footnote{If
we include $m_0$ as a mass insertion in Feynman diagrams, then our
answer is necessarily a series expansion in $m_0$. Using the WKB
approximation, we will be neglecting terms suppressed by higher
powers of $1/(ER)$, but keeping terms to all orders in $m_0$.}. The
wave equation is given by \be \left[\Box + m^2\right] \chi &=& 0,
\ee where the space dependent mass is \be m^2(x_3) &=&
m_0^2(1-B(x_3)),\ee and $B(x_3)$ is a ``bump'' function which
vanishes at infinity and peaks to unity in a compact region of size
$R$. From section \ref{sec:grossprop} we know that $B(r)$ goes as
$1/r$ for large $r$,  but for now lets us take $B(x_3)$ to be a
general function. Consider the following WKB ansatz solution: \be
\label{Definef}
\chi(x) &=& \exp(i \tilde{p}(x) x) \\
\tilde{p}(x) & \equiv & \left( p_0, p_1, p_2, p_3 + \frac{m_0^2
}{2p_3}f(x_3) \right), \ee where $f(x_3)$ is a function that will be
determined by plugging $\chi$ into the wave equation. From now on, a
tilde on a momentum variable will represent a nontrivial space
dependence of this kind.

If we plug this ansatz back into the wave equation, we find \be
\left[ \Box + m^2 \right] \chi &=& \left[ -p^2 + m_0^2 +
m_0^2\left(-B+f+x_3 f'-\frac{i f'}{p_3} - \frac{i x_3
f''}{2p_3}\right) \right] \chi. \ee If the quantity in curved
parentheses vanishes, then $\chi$ is an eigenstate of the wave
equation. In the regime where $p_3 \gg 1/R$, the terms containing
$1/p_3$ are small, since they go as inverse powers of $p_3 R$, so we
drop them. Now it is easy to solve for $f$ in the resulting
differential equation \be -B + f + x_3 f' &=& 0, \ee yielding \be
\tilde{p}(x) &=&\left(p_0,p_1,p_2, p_3 + \frac{m_0^2}{2 p_3 x_3}
\int^{x_3} B(x_3')dx_3'\right). \ee Thus in the WKB approximation,
the $\chi$ propagator is \be \label{eq:chiprop} G_\chi(x,y) = \int
\frac{i}{p^2 - m_0^2} \times e^{i( \tilde{p}(x)x-\tilde{p}(y)y)},
\ee to zeroth order in $1/(E R)$ and all orders in $m_0^2$.

\subsection{Amplitude}

\label{sec:amplitude}

Next, let us consider the amplitude for a $\phi$ particle of
momentum $p$ to decay into two $\chi$ particles of momenta $k$ and
$q$ in the spatially varying background. To do so we write down the
corresponding time ordered 3-point correlator in coordinate space
\be \langle \chi(y) \chi(z) \phi(w) \rangle & \sim& g \int
G_\chi(y-x) G_\chi(z-x) G_\phi(x-w)d^4 x. \ee  Next, we Fourier
transform to the variables $p$, $k$ and $q$.  In accordance with the
LSZ reduction formula, a scattering amplitude corresponds to the
Fourier transform of the appropriate time ordered correlator with
the pole from each external leg stripped off.  Amputating the legs,
we find that the Feynman amplitude is \be \mathcal{M}(k,q,p) &=& g
\times \int d^4 x \, e^{i(\tilde{k}(x)+\tilde{q}(x)-p)x} ,\\&=&
g \times \delta^{(012)}(k+q-p)F(k_3,q_3,p_3) \\
\label{DefineF} F(k_3,q_3,p_3) &=& \int dx_3 \,
e^{i(\tilde{k}_3(x_3) + \tilde{q}_3(x_3) - p_3)x_3}.
\label{eq:Fdef}\ee where $g$ is the dimensionful coupling strength.
Notice that in the $m_0^2 \rightarrow 0$ limit, $F$ reverts to a
full 4-momentum conserving delta function, as expected. Next, let us
evaluate $F$.

By separating a factor of $\exp i (k_3 +q_3 -p_3)x$ in the integrand
of $F$, it is clear that $F$ is simply the Fourier transform of the
quantity \be \exp \left[i m_0^2 \left( \frac{1}{k_3} + \frac{1}{q_3}
\right) \int^{x_3} B(x_3') dx_3' \right] .\ee Let us consider the
case in which the Higgs profile is a square bump function and so $B$ is
unity for $|x_3| < R$ and $0$ otherwise. Given this simplification
the integral is easy to evaluate piece-wise and $F$ takes the simple
form \be F &\sim& \lim_{\epsilon \rightarrow 0} \frac{
\sin(a/\epsilon + R b)}{a} - \frac{b}{a}\frac{ \sin R(a+b)}{(a+b)} ,\\
&\sim& \pi \delta(a) - \frac{b}{a}\frac{ \sin R(a+b)}{(a+b)}, \ee
where we are using $\epsilon$ to regulate the $\delta$ function, and
for convenience we have defined \be a &=& k_3 + q_3 -p_3,\\ b &=&
m_0^2 \left( \frac{1}{k_3} + \frac{1}{q_3} \right). \ee Since $R$ is
the largest length scale in the problem, we can actually simplify
$F$ even further, writing \be F &\sim& \delta(a) - \frac{b}{a}
\delta_R(a+b), \label{eq:niceanswer}\ee where $\delta_R$ denotes an
$R$-regulated delta function with width $1/R$ and height $R$.
Physically, $F$ takes this form because it receives two
contributions, corresponding to exact momentum conservation and
deviations from momentum conservation, set by the scale $1/R$.

Also, note that our results are parametrically correct even if the
Higgs profile differs from the square bump function form which we
have assumed. Looking at other forms for $B$, we find it is only
really necessary that $B \approx 1$ in the region $x_3 \in [-R,R]$.

Next, let us integrate over phase space and compute the total
probability of decay using this amplitude.

\subsection{Probability}

The decay rate for a 1 $\rightarrow$ 2 process is given by \be
\Gamma &\sim& \frac{1}{E}\int \frac{d^3 k d^3 q}{k_0
q_0}|\mathcal{M}|^2. \ee In translationally invariant theories,
$\mathcal{M}$ is proportional to a 4-momentum conserving delta
function, so $|\mathcal{M}|^2$ is a product of squares of delta
functions.  While naively this introduces a divergent $\delta(0)$
term, we normally divide $\Gamma$ by the volume of spacetime, hence
removing these factors.
However, in our case, $F$ has a component that exactly conserves
4-momentum and a component that violates momentum in the $x_3$
direction by an amount $1/R$. From basic kinematics, we know that
the contribution to $\Gamma$ from the 4-momentum conserving piece
will not contribute, since a massless particle cannot decay into two
massive particles in free space.  For this reason, we will only need
to compute the momentum violating contribution to $|F|^2$. Since we
are not dividing by the size of the $x_3$ direction, we are actually
computing the decay rate integrated over all of $x_3$, i.e. the
total probability of decay (see appendix \ref{sec:crosssection}
for details).

For an incoming momentum \be p=(E,0,0,E),\ee the (dimensionless)
decay probability in our toy model is \be P_{\phi \rightarrow
\chi\chi} &\sim & \frac{g^2}{E}\int \frac{d^3 k}{ k_0}\frac{d^3
q}{q_0} \delta(k_0+q_0-E)\delta(k_1 +
q_1)\delta(k_2+q_2)|F(k_3,q_3,E)|^2 ,
\ee where the energies are \be k_0 &=& \sqrt{\vec{k}^2+m_0^2}, \\
q_0 &=& \sqrt{\vec{q}^2+m_0^2}. \ee It is trivial to integrate
over the transverse $q$ momenta, after which we parameterize the two
transverse $k$ momenta in polar coordinates as
\be (k_1,k_2) &=& k_\perp (\cos \theta,\sin\theta) \\
d^3 k &=& k_\perp dk_\perp d \theta dk_3 . \ee From the energy
conservation delta function we obtain the useful expressions \be k_0
= \frac{1}{2E} (E^2+k_3^2 - q_3^2), \\ q_0 = \frac{1}{2E} (E^2+q_3^2
- k_3^2).\ee Applying these formulae, the delta function becomes \be
\d(k_0 + q_0 -E) &=& \frac{1}{k_\perp}\frac{k_0 q_0}{E} \times \d
(k_\perp - K_\perp), \ee where $K_\perp$ is defined as \be K_\perp^2
&=& \frac{1}{4E^2}(k_3+q_3-E)(k_3-q_3-E)(k_3+q_3+E)(k_3-q_3+E)-m^2.
\ee The factor multiplying the delta function cancels with most of
the integral, eliminating all $k_\perp$ dependence except for the
delta function! Note that for a completely 4-momentum conserving
interaction, $k_3 + q_3 = E$, and so $K_\perp^2 = -m^2 < 0$ and the
$k_\perp$ integral yields zero, corresponding to the fact that a
massless particle cannot decay into two massive particles if
4-momentum is conserved. Since we lack momentum conservation in the
3-direction, $K_\perp^2 > 0$ and the $k_\perp$ integral instead
gives unity.  Consequently, we obtain the probability of a $\phi$
particle to decay into two $\chi$ particles \be
\label{PhaseSpaceIntegral} P_{\phi \rightarrow \chi \chi} &\sim&
\frac{g^2}{E^2}\int dk_3 dq_3 |F(k_3,q_3,E)|^2. \ee The domain of
integration is the compact region \be \sqrt{k_3^2+m_0^2}+\sqrt{q_3^2
+ m_0^2} \le E, \ee as derived from energy conservation.  This
inequality is saturated when $k_\perp$ is zero.  Note that we have
made no approximations in evaluating this phase space integral,
so the result, written in terms of $F$, is correct up to numerical
coefficients.

\subsection{From Toy Scalars to Gauge Bosons}

\label{FromToyScalarstoGaugeBosons}

Given $P_{\phi \rightarrow \chi\chi}$, it is straightforward to
obtain $P_{A \rightarrow W W}$, the probability of a photon to decay
to two $W$'s. The only parametric difference in the two calculations
is that in the gauge theory calculation, $g$ becomes a dimensionless
coupling and the three gauge boson interaction has an extra factor
of momentum due to the derivative coupling. This introduces an extra
factor of $E^2$ into the decay probability. Moreover, since the
outgoing $W$'s can be any of $N$ gauge bosons, we also get a factor
of $N$, yielding \be P_{A \rightarrow WW} &\sim& \lambda \int dk_3
dq_3 |F(k_3,q_3,E)|^2.\ee

 Before we plug in for $F$, let us pause to note a high energy subtlety which
was mentioned earlier in terms of soft collinear divergences.
Naively, we would be tempted to simply set $b=-a$ in $F$ because of
the delta function of $a+b$. However, since $b$ is only fixed to be
equal $-a$ up to a $1/R$ width, this replacement is only valid if
$b$ is of order $1/R$ or more, i$.$e$.$ if \be
m_0^2\left(\frac{1}{k_3}+\frac{1}{q_3}\right)
> \frac{1}{R}, \label{k3Constraint}\ee which is precisely the soft collinear bound
derived in Eq.~(\ref{eq:softcoll}).  As we will show, the energy
scaling of the decay probability is quite different above and below
this bound.

First, let us consider the $E < m_0^2 R$ regime.  Setting $b = -a$
and integrating $|F|^2$ over $k_3$ and $q_3$ (ignoring contributions
from the 4-momentum conserving delta function), we find that \be
P_{A\rightarrow W W} & \sim & \lambda \int dk_3
dq_3 \left|\delta_R\left(a+b\right)\right|^2\\
& \sim & \lambda \delta_R(0) \int dk_3,\\ &\sim& \lambda E R, \quad
\textrm{($E<m_0^2 R$)} \ee which is our final answer for the
probability of a photon to decay to two $W$ bosons at energies $E <
m_0^2 R$.

Next, let us determine the probability of decay in the regime $E >
m_0^2 R$, which we will divide into two phenomenologically
distinct regions of phase space.  First, consider the regime where
$k_3$ is less than $m_0^2 R$ but $q_3$ is large enough that their
sum, $E$, exceeds $m_0^2 R$ (obviously our result is symmetric
under $k_3 \leftrightarrow q_3$). We will denote this as the
``Bremsstrahlung'' regime because one of the outgoing particles is
much softer than the other.  Since Eq.~(\ref{k3Constraint})
obtains in this limit, the entire discussion of the previous
paragraphs applies except that the $k_3$ integral is bounded by
$k_3 < m_0^2 R$, and so \be P_{A \rightarrow W W} \sim \lambda
m_0^2 R^2. \quad \textrm{($E>m_0^2 R$, Bremsstrahlung)} \ee If we
are instead interested in a ``symmetric decay,'' defined by $k_3,
q_3
> m_0^2 R$, then it becomes necessary to include the $b/a$ in our
expression for $F$, and the answer becomes \be P_{A \rightarrow W
W} \sim \frac{\lambda m_0^4 R^3}{E}. \quad \textrm{($E>m_0^2 R$,
symmetric decay)} \ee

Note the crucial difference between these two regions of phase
space -- the first corresponds to decay products of energy $m_0^2
R$ and $E - m_0^2 R$ while the second corresponds to two decay
products with energies roughly of order $E$.

\subsection{Higher Order Effects and Approximations}

Thus far we have only taken into account the role of three gauge
boson interactions mediating binary decays in the plasma puddle
atmosphere.  However, it is fair to ask whether these contribution
necessarily dominate over the four gauge boson interactions, which
mediate trinary decays.  In this section compare the relative sizes
of $P_{\rm A \rightarrow W W W}$ and $P_{\rm A \rightarrow W W
\rightarrow W W W}$, and argue that the former contribution is
subdominant.

Repeating our procedure from section \ref{sec:amplitude}, we find
the contribution to the decay amplitude from the four gauge boson
interaction, \be \mathcal{M}(k,q,r,p) &=& g^2 \times \int d^4 x \,
e^{i(\tilde{k}(x)+\tilde{q}(x)+\tilde{r}(x)-p)x} .\ee  Again, this
amplitude simplifies to the form $\delta_R(a+b)$, where this time
\be a &=& k_3 + q_3 + r_3 - p_3 ,
\\ b &=& m_0^2
\left(\frac{1}{k_3}+\frac{1}{q_3}+\frac{1}{r_3}\right).\ee  Summing
over $N^2$ possible final states, we find that \be P_{A\rightarrow
WWW} &\sim& \lambda^2 E R.\ee  Despite multiple phase space
integrals, the form of this answer is expected because $|F|^2 \sim
|\delta_R(a+b)|^2$ yields precisely one factor of $R$, and factors
of $E$ make up the rest of the expression.

Comparing, we see that the probability of one trinary decay
$P_{A\rightarrow WWW}$ is much less than that of two consecutive
binary decays \be P_{A\rightarrow WW \rightarrow WWW} &\sim&
(\lambda E R)^2,\ee so it is the latter that dominates the 1
$\rightarrow$ 3 decay.  Thus, the dominant contribution to the
decay is binary and our expression from section
\ref{sec:absorption} is valid.  Similarly, loop effects are not
large even though the $1 \rightarrow n$ rate is large at high
energies, because only initial and final states involve
propagators that are almost on-shell.

\section{Rates and Cross Sections}

\label{sec:crosssection}

Normally, we are interested quantities such as the cross section and
decay rate. However, in our setup the ``decay rate'' is position
dependent, and so to avoid this complication we simply compute the
total probability that the incident particle will decay. This
requires a slight revision of the standard formulas for converting
S-matrix elements into physical observables.

Let us begin with Weinberg's \cite{Weinberg} formula Eq.~$3.4.9$.
Putting the universe in a box of size $L$ and in a time interval
of size $T$, we have that \be dP(\alpha \rightarrow \beta) \sim
\frac{1}{L^3} |S_{\beta \alpha} |^2 d \beta. \ee  In a 4-momentum
conserving theory, the S-matrix element is defined as \be S_{\beta
\alpha} \sim i \delta^{0123}(p_\beta - p_\alpha) M_{\beta \alpha},
\ee where we identify $\delta(0) \sim L$.  However, in our case
momentum is violated in the $x_3$ direction, so we have that \be
S_{\beta \alpha} \sim i \delta^{012}(p_\beta - p_\alpha)
F(p_{\beta 3} - p_{\alpha 3}) . \ee Thus the cancellation of
volume factors will not be complete.  Instead \be dP(\alpha
\rightarrow \beta) \sim \frac{T}{L} |M_{\beta \alpha}|^2
\delta^{012}(p_\beta - p_\alpha) \left| F(p_{\beta 3} - p_{\alpha
3}) \right|^2 d \beta. \ee Normally we would divide by $T$ and
take the $L\rightarrow \infty$ limit, but in our case this sends
the probability to zero.  This is just the statement that a
particle in infinite volume will, on average, take an infinite
time to hit the domain wall. Similarly, given a finite volume, if
$T \rightarrow \infty$ then probability will blow up.

Thus we should take $T, L \rightarrow \infty$ with $T/L$ held
constant. It seems that this leads to an arbitrary factor, but
physically this factor should be one, since it just corresponds to
the number of times the particle crosses the domain wall.

\end{document}